\documentclass[12pt]{article}
\pdfoutput=1
\usepackage{epsfig}
\usepackage{amssymb}
\usepackage{amsfonts}
\usepackage{subfigure}
\def\beq{\begin{equation}}
\def\eeq{\end{equation}}
\def\bea{\begin{eqnarray}}
\def\eea{\end{eqnarray}}
\def\ksl{\hbox{\hbox{${k}$}}\kern-1.9mm{\hbox{${/}$}}}

\newcommand{\nn}{\nonumber}
\newcommand{\text}{\rm}

\newcommand{\GeV}{\textrm{GeV}}

\def\lsim{\raise0.3ex\hbox{$\;<$\kern-0.75em\raise-1.1ex\hbox{$\sim\;$}}} 

\def\gsim{\raise0.3ex\hbox{$\;>$\kern-0.75em\raise-1.1ex\hbox{$\sim\;$}}}

\begin{document}

\begin{center} 
{\bf \large  Vacuum Stability in $U(1)$-Prime Extensions of the Standard Model \\
with TeV Scale Right Handed Neutrinos}
 \vspace{0.2cm}
{\bf \Large } 

\vspace{1.5cm}
{\bf Claudio Corian\`{o}, Luigi Delle Rose and Carlo Marzo}
\vspace{1cm}

{\it Dipartimento di Matematica e Fisica "Ennio De Giorgi", 
Universit\`{a} del Salento and \\ INFN-Lecce, Via Arnesano, 73100 Lecce, Italy\footnote{claudio.coriano@le.infn.it, luigi.dellerose@le.infn.it, carlo.marzo@le.infn.it}}\\

\vspace{.5cm}
\begin{abstract} 
We investigate a minimal $U(1)'$ extension of the Standard Model with one extra complex scalar and generic gauge charge assignments. We use a type-I seesaw mechanism with three heavy right handed neutrinos to illustrate the constraints on the charges, on their mass and on the mixing angle of the two scalars, derived by requiring the vacuum stability of the scalar potential. 
We focus our study on a scenario which could be accessible at the LHC, by selecting a vacuum expectation value of the extra Higgs in the TeV range and determining the constraints that emerge in the parameter space.  To illustrate the generality of the approach, specific gauge choices corresponding to $U(1)_{B-L}, U(1)_R$ and $U(1)_\chi$ are separately analyzed. Our results are based on a modified expression of one of the $\beta$ functions of the quartic couplings of the scalar potential compared to the previous literature. This is due to a change in the coefficient of the Yukawa term of the right handed neutrinos. Differently from previous analysis, we show that this coupling may destabilize the vacuum.

\end{abstract}
\end{center}
\newpage
\section{Introduction} 
In the Standard Model (SM), it has been observed since long ago \cite{Cabibbo:1979ay, Lindner:1985uk, Lindner:1988ww, Ford:1992mv} that the requirement of vacuum stability up to the unification scale and beyond, and the absence of a Landau pole under the renormalization group (RG) evolution, constrain the value of the Higgs mass $(m_h)$ and the size of the Yukawa couplings of the heavy fermions \cite{Bezrukov:2012sa,Buttazzo:2013uya}. Lower and upper bounds on $m_h$ have been derived and shown to depend more or less significantly on the size of $Y_t$, the Yukawa of the top quark,  which can drive the quartic Higgs coupling to become negative beyond a certain scale. This situation can be ameliorated with the addition of extra scalars, either in the form of SM singlets, in some cases even taking the role of dark matter components \cite{Kadastik:2011aa,Chen:2012faa}, or by a modification of the scalar potential. Crucial in this type of analysis is the sign and the size of the various contributions to the $\beta_\lambda$ function of the quartic Higgs coupling $(\lambda)$, which is negative for fermions and positive for scalars. At the same time, the size of the same coupling at the electroweak scale $(v)$, i.e. at the starting scale of the evolution, turns out to be of extreme importance in driving $\lambda$ either towards a non-perturbative region or to render the Higgs potential unstable in the far ultraviolet.  

In this work we are going to investigate the constraints imposed by the condition of vacuum stability in a rather minimal extension of the SM enlarged by an extra $U(1)'$ symmetry and one extra Higgs scalar. This is a SM singlet which triggers the spontaneous breaking of the extra abelian symmetry with a vev $v'$ assumed to lay around the TeV scale. Obviously, this specific choice selects an interesting subregion of parameter space with a heavy Higgs in the 
TeV range which could be explored at the LHC in the near future. 
In particular, we are going to examine how these constraints are modified by the inclusion of three right handed neutrinos, taken to be 
SM singlets.  We require the mass of the SM neutrinos to be generated  by a type-I seesaw mechanism \cite{GellMann:1980vs, Mohapatra:1979ia, Yanagida:1979as}, with a Majorana mass scale chosen in the TeV region and a Yukawa coupling $Y_\nu$ of the three SM neutrinos of 
$\lesssim 10^{-6}$. Our results differ from previous interesting analysis of a similar model \cite{Datta:2013mta}, investigated in the specific case of a $U(1)_{B-L}$ symmetry, being based on a recalculated expression of one of the $\beta$ functions of the quartic couplings of the scalar potential, as specified below. We will see that the requirement of vacuum stability under the evolution sets significant constraints on the mass of the right handed neutrinos. The result will also depend on the mixing angle $\theta$ between the heavy and the light Higgs and on the mass of the heaviest scalar $m_{h_2}$. 
\section{The Model} 
The model that we consider has the family structure of the SM with three generations, and a gauge symmetry of the 
form $SU(3)_c \times SU(2)_w \times U(1)_Y \times U(1)'$, which is the SM gauge group enlarged by a single extra $U(1)'$. We allow a kinetic mixing between the two $U(1)$ abelian fields, which can be reabsorbed in the definition of the gauge covariant derivative as
\bea
\mathcal D_\mu = \partial_\mu + i g_3 T^a G^a_\mu + i g_2 t^a W^a_\mu + i g Y B_\mu + i (\tilde g Y +  g' Y') B'_\mu \,,
\eea
where $g$ and $g'$ are the coupling constants associated with $U(1)_Y$ and $U(1)'$ respectively and $Y$ and $Y'$ are the corresponding charges. The coupling $\tilde g$ describes the mixing.

Our choice for the scalar sector is deliberately minimal, characterized by the usual SM  $SU(2)_w$ Higgs doublet $H$ enlarged just by an extra SM singlet complex scalar $\phi$. In this case, the most general renormalizable scalar potential is given by
\bea
V(H,\phi) = m_1^2 H^\dag H + m_2^2 \phi^\dag \phi + \lambda_1 (H^\dag H)^2 + \lambda_2  (\phi^\dag \phi)^2 + \lambda_3  (H^\dag H) ( \phi^\dag \phi) \,
\eea
constrained by the following conditions on its quartic couplings
\bea
\label{stabilitycond}
\lambda_1 > 0\,, \quad \lambda_2 >0 \,, \quad 4 \lambda_1 \lambda_2 - \lambda_3^2 > 0 \,,
\eea
in order to ensure its stability. The ground state of the theory is characterized by the vacuum expectation values (vev) of the doublet $H$ and of the singlet $\phi$ fields
\bea
< H > = \frac{1}{\sqrt{2}} \left( \begin{tabular}{c} 0 \\ $v$ \end{tabular} \right) \,, \qquad <\phi> = \frac{v'}{\sqrt{2}}\,,
\eea
whose expressions, determined by the minimization conditions, take the form
\bea
v^2 = \frac{ m_2^2 \, \lambda_3/2 - m_1^2 \, \lambda_2 }{\lambda_1 \lambda_2 - \lambda_3^2 /4} \,, \qquad
v'^2 = \frac{ m_1^2 \, \lambda_3/2 - m_2^2 \, \lambda_1 }{\lambda_1 \lambda_2 - \lambda_3^2 /4} \,.
\eea
After spontaneous symmetry breaking, the mixing between the two scalar fields can be removed by a rotation into the two mass eigenstates $h_1$ and $h_2$
\bea
\left( \begin{array}{c} h_1 \\ h_2 \end{array} \right) = \left( \begin{array}{cc} \cos \theta & - \sin \theta \\  \sin \theta & \cos \theta \end{array} \right)  \left( \begin{array}{c} H  \\ \phi \end{array} \right)
\eea 
where the mixing angle $\theta$ is given by
\bea
\tan 2 \theta = \frac{\lambda_3 v v'}{\lambda_1 v^2 - \lambda_2 v'^2} \,,
\eea
with $- \pi/2 < \theta < \pi/2$. Their masses are given by
\bea
m_{h_{1,2}}^2 = \lambda_1 v^2 + \lambda_2 v'^2 \mp \sqrt{\left( \lambda_1 v^2 - \lambda_2 v'^2\right)^2 + \left( \lambda_3 v v' \right)^2} \,,
\eea
and the three quartic couplings are expressed in terms of these as 
\bea
\label{lambdas}
\lambda_1 &=& \frac{m_{h_1}^2}{4v^2}(1+\cos 2\theta) + \frac{m_{h_2}^2}{4 v^2}(1-\cos 2 \theta) \,, \nn \\
\lambda_2 &=& \frac{m_{h_1}^2}{4v'^2}(1-\cos 2\theta) + \frac{m_{h_2}^2}{4 v'^2}(1+\cos 2 \theta) \,, \nn \\
\lambda_3 &=& \sin 2 \theta \left( \frac{m_{h_2}^2 - m_{h_1}^2}{2 v v'}\right) \,,
\eea
which can be used to set the initial conditions on the scalar couplings through the physical masses $m_{h_{1,2}}$, the two vevs $v, v'$ and the mixing angle $\theta$. 

\begin{table}
\centering
\begin{tabular}{|c|c|c|c|c|}
\hline
    & $SU(3)_c$ & $SU(2)_w$ & $U(1)_Y$ & $U(1)'$ \\ 
\hline 
$Q_L$ & 3 & 2 & 1/6 & $z_Q$ \\
$u_R$ & 3 & 1 & 2/3 & $z_u$ \\
$d_R$ & 3 & 1 & -1/3 & $ 2z_Q-z_u$ \\
$L$ & 1 & 2 & -1/2 & $-3z_Q$ \\
$e_R$ & 1 & 1 & -1 & $-2z_Q - z_u$ \\
$H$ & 1 & 2 & 1/2 & $z_H$ \\
$\nu_{R, k}$ & 1 & 1 & 0 & $z_k$ \\
$\phi$ & 1 & 1 & 0 & $z_\phi$ \\ \hline
\end{tabular}
\caption{Charge assignment of fermions and scalars in the $U(1)'$ SM extension. \label{Table1}} 
\end{table}
\begin{table}
\centering
\begin{tabular}{|c|c|c|c|c|c|c|c|c|}
\hline
                          & $Q_L$ & $u_R$ & $d_R$ & $L$ & $e_R$  & $\nu_R$ & $H$ & $\phi$ \\
\hline
$U(1)_{B-L}$ & 1/3 & 1/3 & 1/3 & -1 & -1 & -1 & 0 & 2 \\
$U(1)_{R}$ & 0 & -1 & 1 & 0 & 1 & -1 & -1 &  2 \\
$U(1)_{\chi}$ & 1/5 & -1/5 & 3/5 & -3/5& -1/5& -1& -2/5& 2  \\
\hline
\end{tabular}
\caption{Specific charge assignments in the $U(1)_{B-L}$, $U(1)_{R}$ and $U(1)_{\chi}$ models. \label{Table2}} 
\end{table}

We show in Tab.~\ref{Table1} the charge assignments of a general non-exotic $U(1)'$ extension \cite{Appelquist:2002mw}. The fermion charges are obtained by imposing the cancellation of all the anomalies, including the gravitational ones. We assume the charges of the $U(1)'$ to be family universal, with the equations for the gravitational anomalies imposed symmetrically respect to the three families, which sets the charges of the three right handed neutrinos to be equal ($z_k\equiv z_\nu,\, k=1,2,3$). The solutions of the anomaly equations are parameterized by two free $U(1)'$ charges, $z_Q$ and $z_u$, of the left-handed quark doublet $Q_L$ and of the right handed up quark $u_R$. 
Notice that the generators of the extra abelian symmetry can be re-expressed, in general, as a linear combination of the SM hypercharge, $Y$, and the $B-L$ quantum number, $Y_{B-L}$, 
\beq
\label{comb}
Y'=\alpha_Y Y +\alpha_{B-L} Y_{B-L}, 
\eeq
where we have denoted with $B$ and $L$ the baryon and lepton numbers respectively. In Eq.~(\ref{comb}) the coefficients $\alpha_Y$ and $\alpha_{B-L}$ are functions of the set of the independent charges of each model realization, as determined by the conditions of anomaly cancellations, and are explicitly given by 
\beq
\label{alpha}
\alpha_Y = 2 z_u - 2 z_Q,  \qquad \qquad\alpha_{B-L} = 4 z_Q- z_u.
\eeq
Concerning all those charges not constrained by the anomaly cancellation, we can use the $U(1)'$ gauge invariance of the Yukawa Lagrangian to fix them. In particular, the scalar doublet charge $z_H$ is fixed by the SM Yukawa interactions to $z_H = z_u - z_Q$.
Other constraints on $z_k$ and $z_\phi$ can be imposed from additional Yukawa terms which are introduced to implement the type-I seesaw mechanism and, therefore, play an important role in the generation of small neutrino masses for the SM neutrinos. Notice that we will consider only dimension-4 operators, namely the Yukawa interactions which generate, through spontaneous symmetry breaking, a Dirac mass term for the SM neutrinos and a Majorana mass for the right handed ones
\bea
\mathcal L_{yuk} &=&  \mathcal L_{SM\,yuk} - Y_\nu \, L \cdot H \nu_R^c - Y_N \, \phi \, \nu_R \nu_R + h.c. \,.
\eea
The requirement of their gauge invariance fixes the remaining charges to $z_\nu = - 4 z_Q + z_u$ and $z_\phi = - 2 z_\nu$. \\
For definiteness, in the following, we will also consider three particular charge assignments, corresponding to the models $U(1)_{B-L}$, 
$U(1)_R$ and $U(1)_\chi$, obtained as special cases of the general assignments given in Tab.~\ref{Table1}. These are given in in Tab.~\ref{Table2}. The $U(1)_\chi$ can emerge, for instance, from the $SO(10)$ grand unified theory (GUT) via $SO(10)\to SU(5)\times U(1)_\chi$.
We will consider these charge assignments simply as specific realizations of the extra abelian symmetry, stressing, for the rest, only on the general features that emerge from the requirement of vacuum stability in these models, with no reference to their GUT origin.

After spontaneous symmetry breaking, the effective Lagrangian describing the neutrino masses will contain Dirac $(M_d)$ and Majorana ($M_m$) mass terms of the form 
\bea \label{TypeISeesawEffective}
\mathcal L_{m_\nu} &=&   -    \nu_L  M_{d} \, \nu_R^c - \frac{1}{2}  \nu_R M_{m} \nu_R + \textrm{h.c.}  \, ,
\eea
where the mass matrices
\bea
M_{d} =  Y_\nu \, \frac{v}{\sqrt{2}} \,, \qquad
M_{m} = \sqrt{2} \, Y_N \, v' 
\eea
inherit the flavour index structure from the corresponding Yukawa ones. As a result of the seesaw mechanism, the mass of the heavy neutrino is of the order of the Majorana mass ($m_{\nu_h}\sim M_m$) while the mass of the SM neutrinos ($m_\nu \sim 1$ eV) is given by the relation 
\beq
m_\nu \sim  \frac{1}{2 \sqrt{2}}\frac{Y_\nu^2 v^2}{Y_N v'}.
\eeq
Being interested in a vev $v'$ of the order of the TeV, the Yukawa $Y_\nu$ must be $\lesssim 10^{-6}$, which is essential to reproduce the light neutrino masses, and, therefore, can be neglected in the RG evolution. On the other hand, $Y_N$, the Yukawa of the heavy right handed neutrinos, could be even of $O(1)$ and, henceforth, it plays an important role. 

All the couplings of the Lagrangian evolve with RG equations whose general expressions are too lengthy to be given here. We just report the expressions of the one-loop $\beta$ functions related to the parameters $\lambda_i$, for the simpler case of $U(1)_{B-L}$, having retained only the top quark and the right handed neutrino contributions in the fermion sector. They take the form
\bea
\label{betafuncs}
\beta_{\lambda_1} &=& 
 24 \lambda _1^2+\lambda _3^2 + \lambda _1 \left(12
   Y_t^2  -3 g^2  -  3 \tilde{g}^2-9 g_2^2 \right) -6 Y_t^4  
+\frac{3 g^4}{8}+\frac{3}{4} g^2 \tilde{g}^2+\frac{3}{4} g^2 g_2^2 \nn \\
&+& \frac{3 \tilde{g}^4}{8}+\frac{3}{4} g_2^2 \tilde{g}^2+\frac{9 g_2^4}{8} \,, \nn \\
\beta_{\lambda_2} &=&
8 \lambda _2  \textrm{tr} (Y_N^2) - 48  \lambda_2 \, g^{'\,2}  - 16
   \textrm{tr} (Y_N^4)+96 g^{'\,4} + 20 \lambda _2^2+2 \lambda _3^2 \,, \nn \\
\beta_{\lambda_3} &=&
 4
   \lambda _3^2+12 \lambda _1 \lambda _3+8 \lambda _2 \lambda _3 + 
\lambda _3 \left\{4 \textrm{tr}\left(Y_N^2\right) + 6 Y_t^2 -\frac{3 g^2}{2}  -\frac{3
   \tilde{g}^2}{2}-\frac{9 g_2^2}{2}- 24 g^{'\,2} \right\}
  +12 g'^2 \tilde{g}^2 \,. \nn \\
\eea
Notice that $\beta_{\lambda_2}$ differs from the expression given in the previous literature in regards to the coefficient 
in front of $Y_N^4$ (see for instance \cite{Datta:2013mta}). This change impacts considerably the RG running of $\lambda_2$, which, for certain values of $Y_N$, does not stay positive along the entire evolution, as found in previous studies, compromising the vacuum stability requirements given in Eq.~(\ref{stabilitycond}).

\section{Numerical Results}
\subsection{Weakly coupled evolution}
In support of a perturbative picture, based on a weak coupling expansion, we start our analysis by demanding that the new gauge coupling constants, $g'$ and $\tilde g$, introduced in the abelian extension, remains less than $\sqrt{4\pi}$, up to some scale $Q$. 
Indeed, the parameters upon which the perturbative expansions are performed are usually of the form $\sqrt{\alpha} = g/\sqrt{4 \pi}$, rather then $g$.
This requirement gives 
\bea
\label{pertconds}
g'(Q') < \sqrt{4\pi}\,, \quad \tilde{g}(Q') < \sqrt{4\pi} \qquad Q' \le Q,
\eea
with the initial conditions at the electroweak scale given by $g'(Q_{ew}) = g'$ and $\tilde{g}(Q_{ew}) = 0$, where $g'$ is a free parameter and we have chosen $Q_{ew} \equiv m_{t}$. In the following we will always assume the vanishing of the abelian kinetic mixing $\tilde{g}$ at the electroweak scale, which is, however, reintroduced by the RG evolution at higher scales. 
\begin{figure}[t]
\centering
\subfigure[]{\includegraphics[scale=0.466]{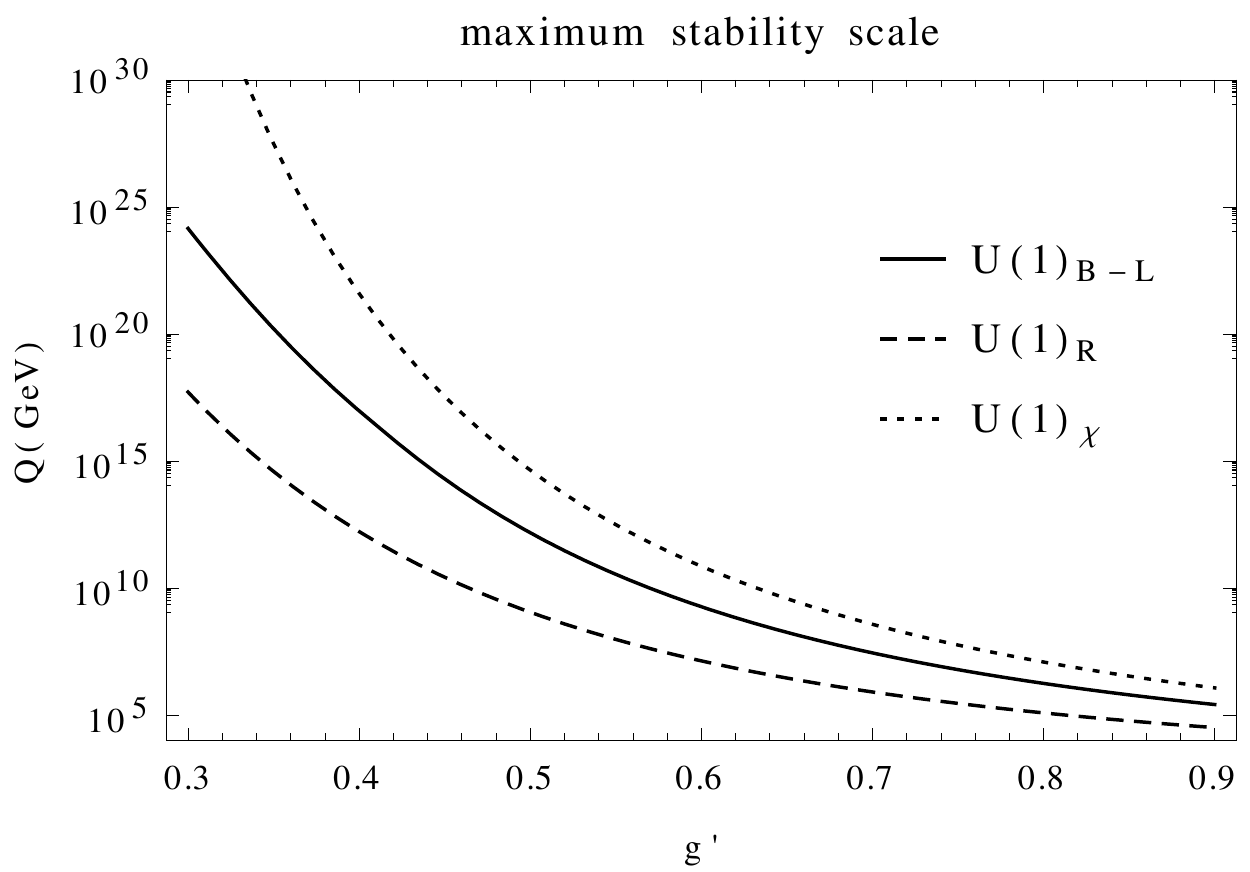}} 
\subfigure[]{\includegraphics[scale=0.466]{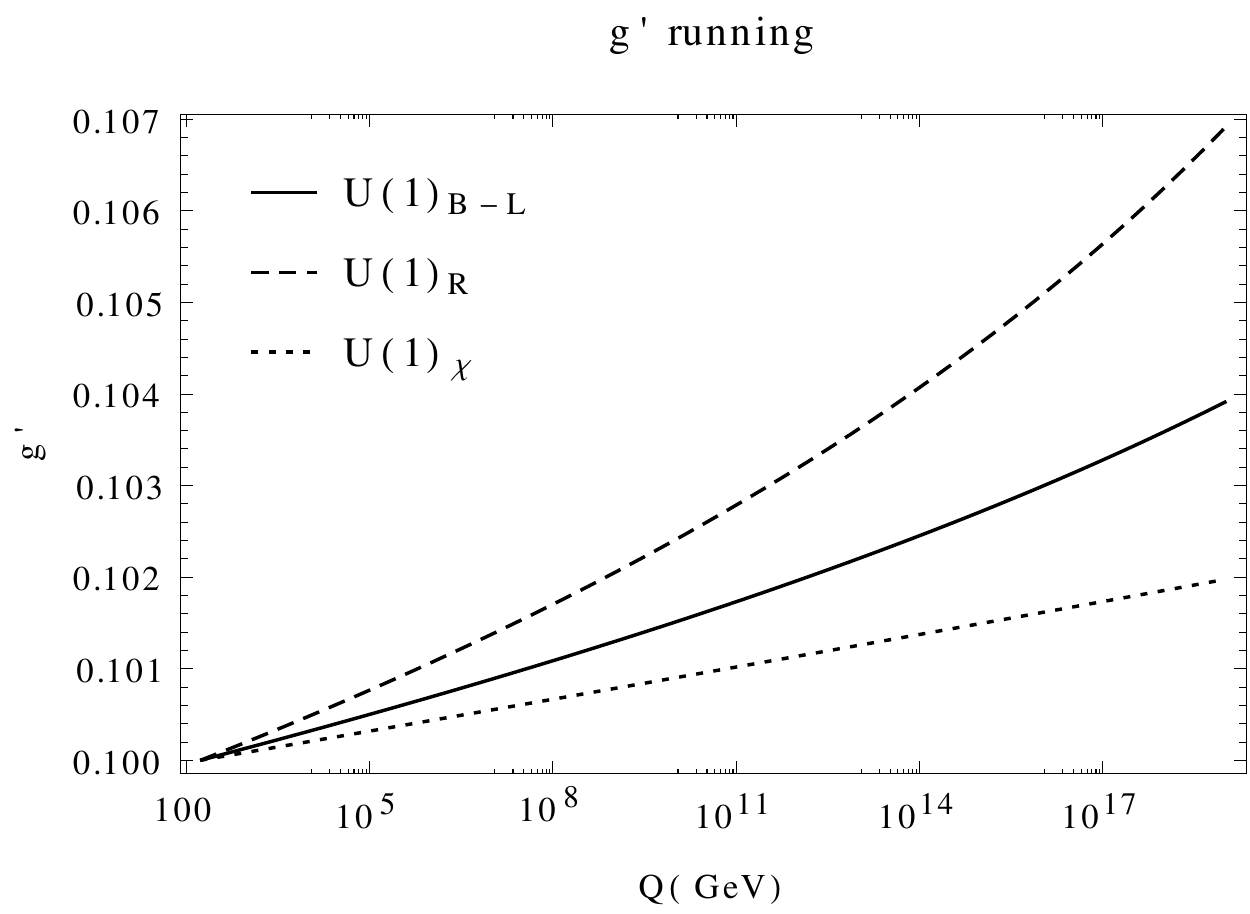}}
\subfigure[]{\includegraphics[scale=0.466]{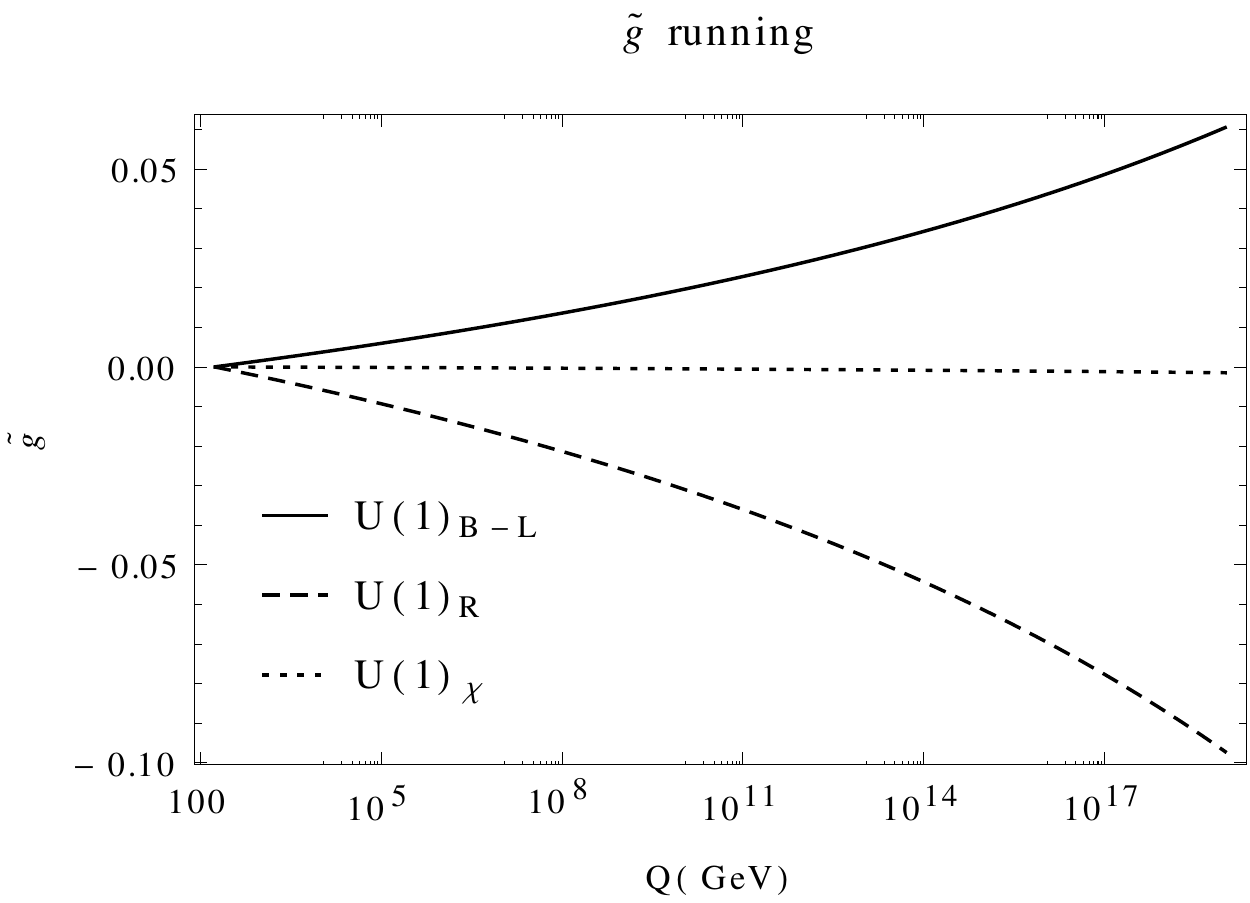}}
\caption{(a) Maximum scale up to which the abelian gauge sector remains perturbative, as a function of the initial condition on $g'$ at the electroweak scale. The gauge mixing coupling $\tilde g$ is assumed to vanish at the same scale. (b) and (c) Running of the $g'$ and $\tilde{g}$ couplings with initial conditions $g'=0.1$ and $\tilde{g} = 0$ at the electroweak scale. \label{Fig.PertCoup}}
\end{figure}
In Fig.~\ref{Fig.PertCoup} (a) we have shown the maximum scale $Q$ up to which the perturbative regime in the abelian sector is maintained as a function of the initial condition $g'$. Results are shown for the three $U(1)'$ extensions discussed above, $U(1)_{B-L}, U(1)_R, U(1)_{\chi}$. For initial conditions $g' < 0.3$, the plots show that a perturbative evolution is allowed up to the Planck scale for all the three models. A more sizeable value of $g'$ at the electroweak scale shows that the evolution violates the weak coupling conditions already at a scale of $10^6-10^7$ GeV, questioning the use of a perturbative expansion beyond such a scale. This trend is quite different for the three models, with the $U(1)_R$ case more significantly affected by the large growth of the coupling and the $U(1)_\chi$ case the least. In the case of $U(1)_{B-L}$, the weak coupling condition is well respected in this model beyond the Planck scale, for initial conditions on the coupling up to $g'\sim 0.35$. 

In Fig. \ref{Fig.PertCoup} (b) we assume the initial conditions $g'=0.1$ and $\tilde g = 0$, which guarantee a perturbative evolution 
up and beyond the Planck scale for the three models, and investigate the changes induced by the evolution on the $g'$ coupling. Up to a large scale of $10^{19}$ GeV these are found to be tiny, at the level of few per mille, showing that for this choice of initial condition, they are essentially frozen. The running of $\tilde{g}$, which quantifies the impact of the kinetic mixing on the evolution, is shown in Fig. \ref{Fig.PertCoup} ({c}), assuming its vanishing at the electroweak scale. It grows/decreases rather modestly in the case of $U(1)_{B-L}$ and of $U(1)_R$ respectively, and stays at zero in the case of $U(1)_\chi$ over the entire evolution, as it should be, being the $SO(10)$-inspired $U(1)_\chi$ model the only orthogonal $U(1)'$ extension of the SM.

\subsection{Weak coupling for a general $U(1)'$}
\begin{figure}[t]
\subfigure[]{\includegraphics[scale=0.45]{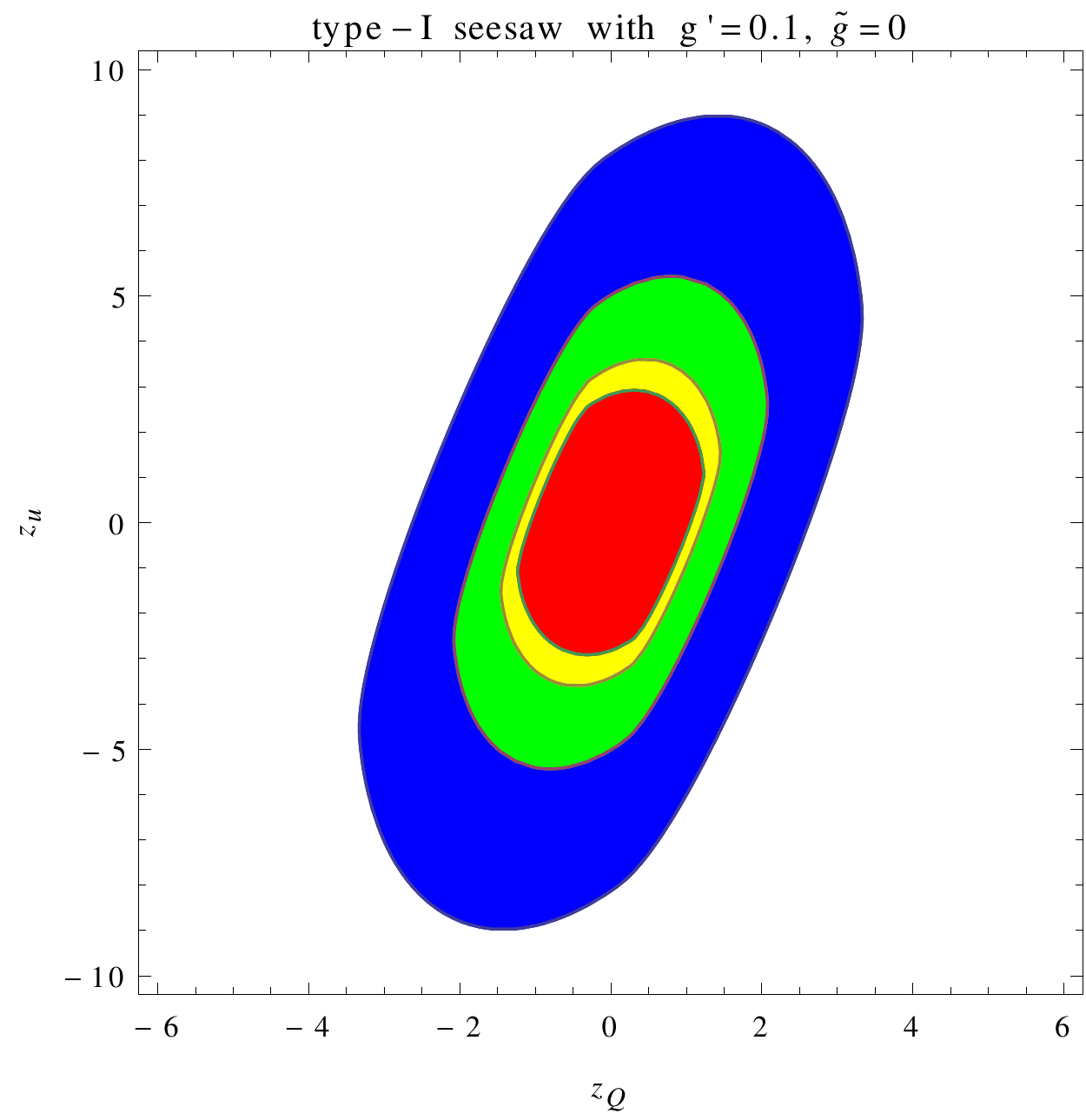}}
\subfigure[]{\includegraphics[scale=0.45]{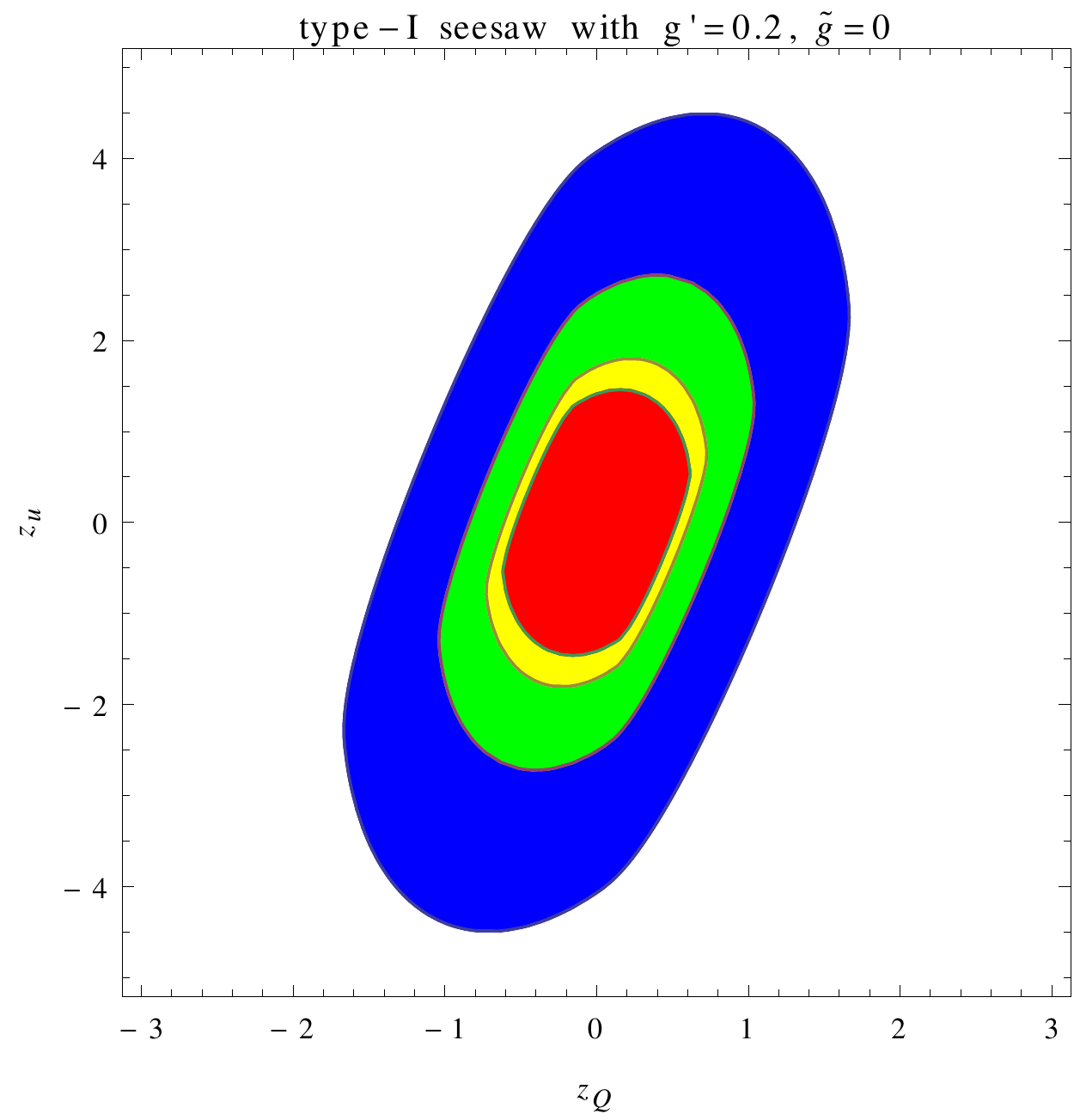}}
\subfigure[]{\includegraphics[scale=0.46]{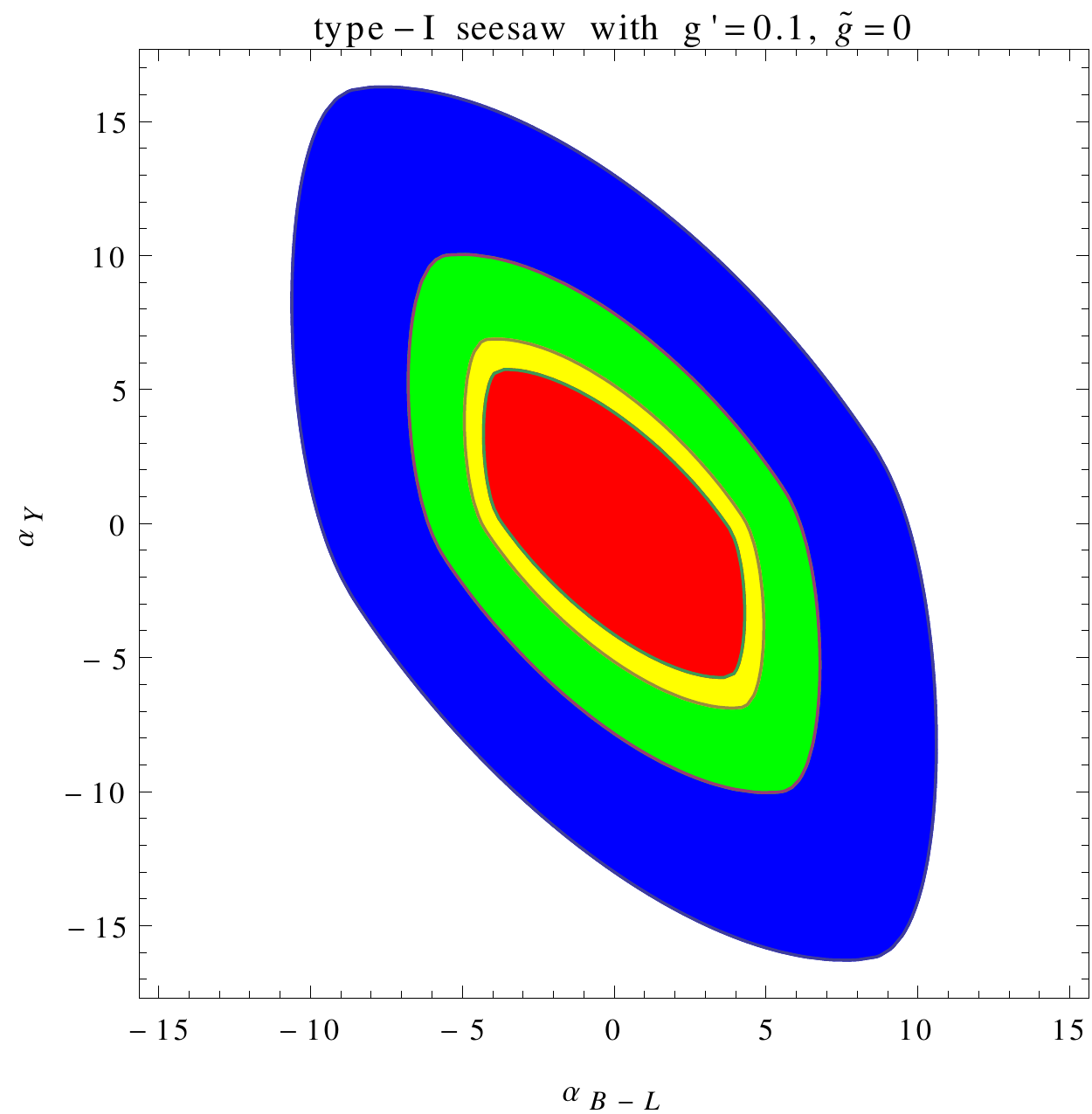}}
\caption{Allowed values of the $U(1)'$ charges, $z_Q$ and $z_u$, for which the perturbative regime is preserved up to $10^5\, \GeV$ (blue region), $10^{9} \, \GeV$ (green region), $10^{15} \, \GeV$ (yellow region) and $10^{19} \, \GeV$ (red region). The first two plots refer to the type-I seesaw scenario for, respectively, $g'=0.1$ and $g'=0.2$ at the electroweak scale. The last one is given in terms of two combinations $\alpha_Y = 2z_u-2z_Q$ and $\alpha_{B-L}  = 4z_Q-z_u$ for $g'=0.1$. \label{Fig.PertCoupss1}}
\end{figure}

It is also interesting to analyze the effects of the charge assignments on the validity of perturbation theory in a general abelian gauge sector, extending the result discussed above. Indeed, choosing a reference value of the initial condition on $g'$, one can repeat the previous analysis, and investigate the weakly coupled region of the theory as a function of the two free $U(1)'$ charges $z_Q$ and $z_u$, here taken as continuous parameters. In Fig.~\ref{Fig.PertCoupss1} we show the region in parameter space of the two independent charges in which perturbation theory is maintained up to $10^5\, \GeV$ (blue region), $10^{9} \, \GeV$ (green region), $10^{15} \, \GeV$ (yellow region) and $10^{19} \, \GeV$ (red region) for two different initial values ($g' = 0.1$ and $g' = 0.2$) at the electroweak scale. The left and the central panel show that the weak coupling expansion up to the Planck scale is tightly bound by charge values $|z_Q| \lesssim 1.5$ and $|z_u| \lesssim 3$. At the same time, the parameter region where the weak coupling conditions are preserved becomes narrower as the initial conditions on the coupling grows. We show in the 
right panel a plot of the same region in the variables $\alpha_Y$ and $\alpha_{B-L}$ as defined by Eq.~(\ref{alpha}). Notice from this last plot that the $U(1)_{B-L}$ projection, obtained for $\alpha_Y=0$, covers the central (red) region characterized by the highest weak coupling scale with $|\alpha_{B-L}|< 5.$

\begin{figure}[t]
\centering
\subfigure[]{\includegraphics[scale=0.45]{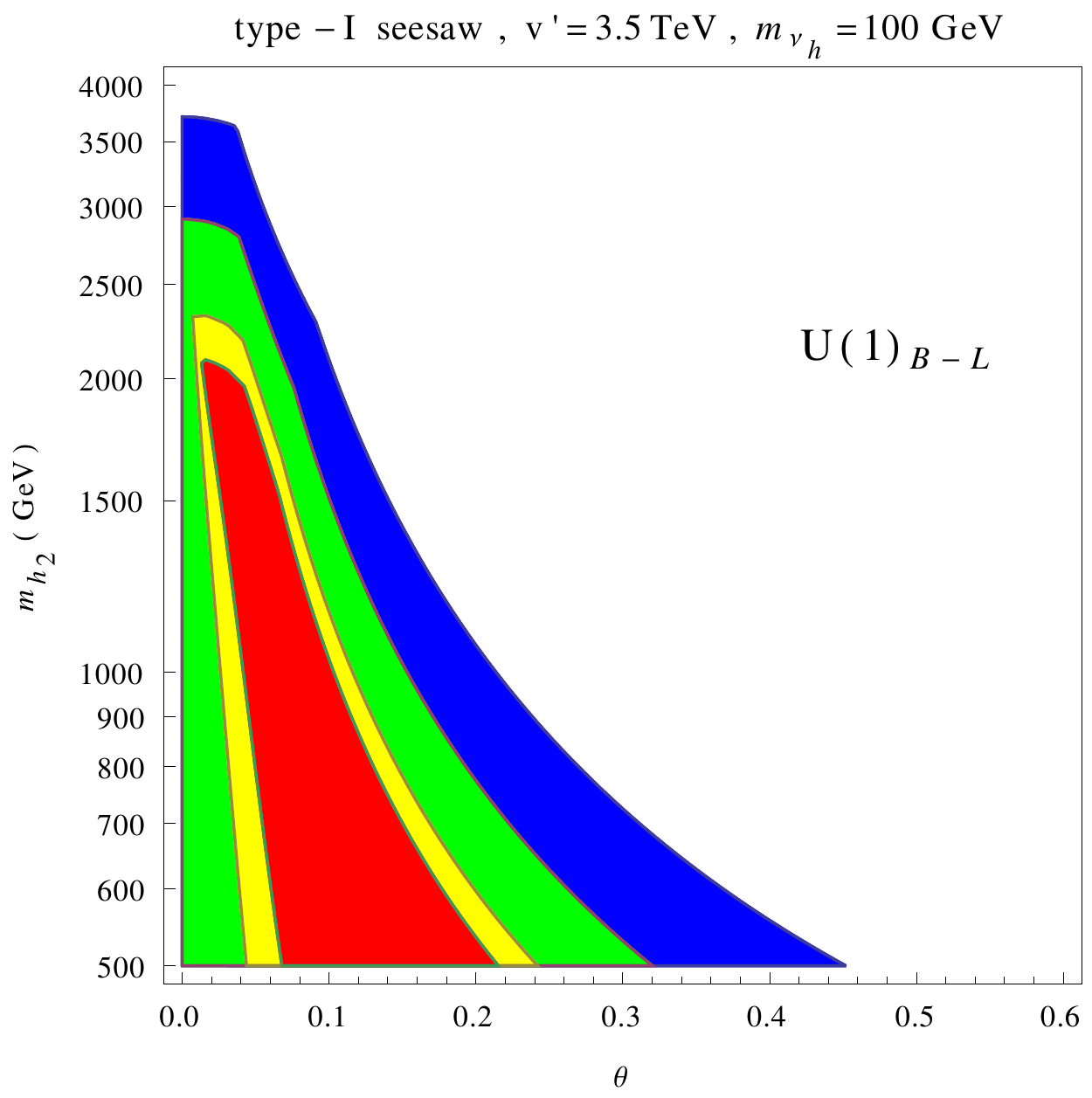}} 
\subfigure[]{\includegraphics[scale=0.45]{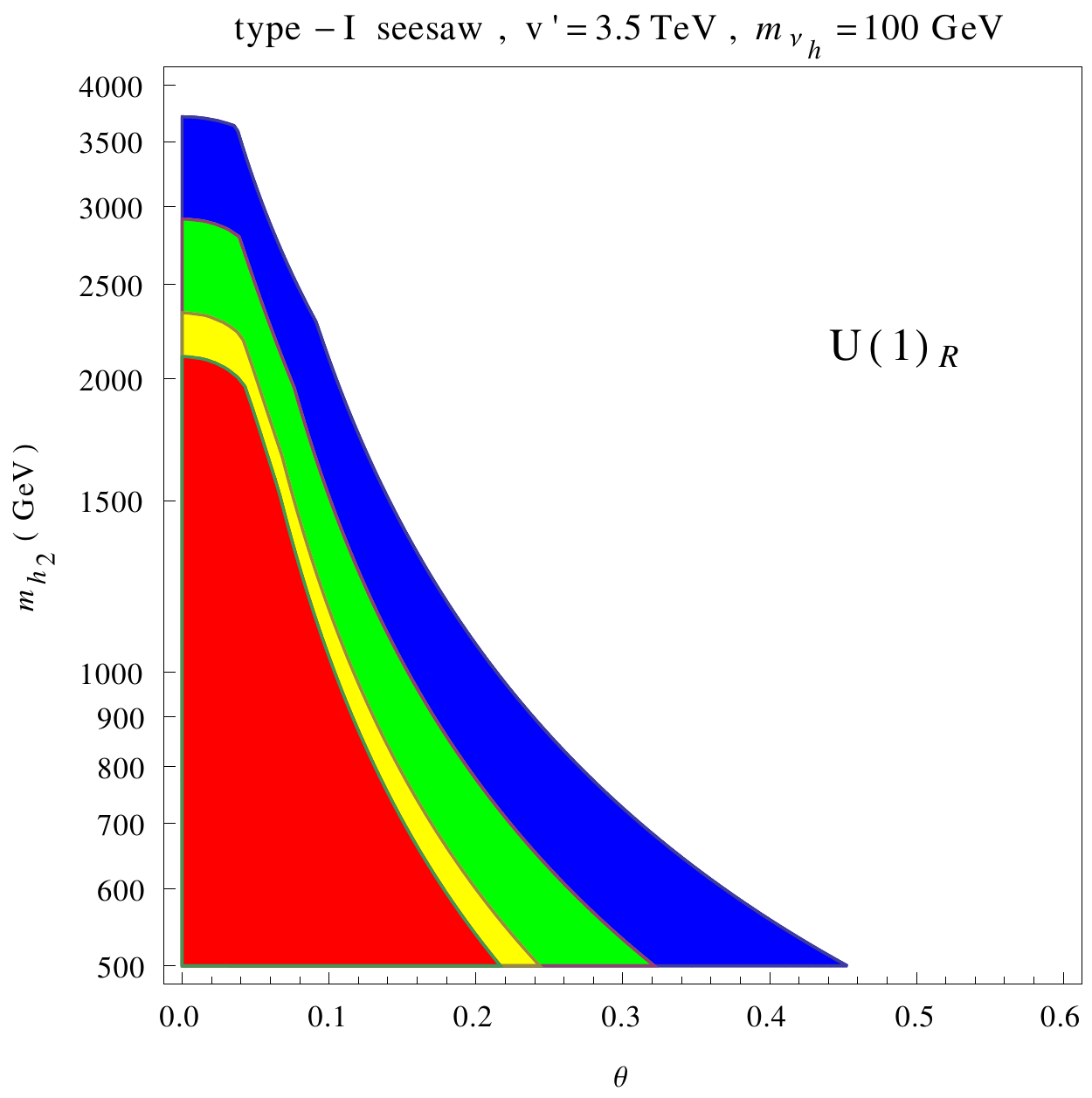}} 
\subfigure[]{\includegraphics[scale=0.45]{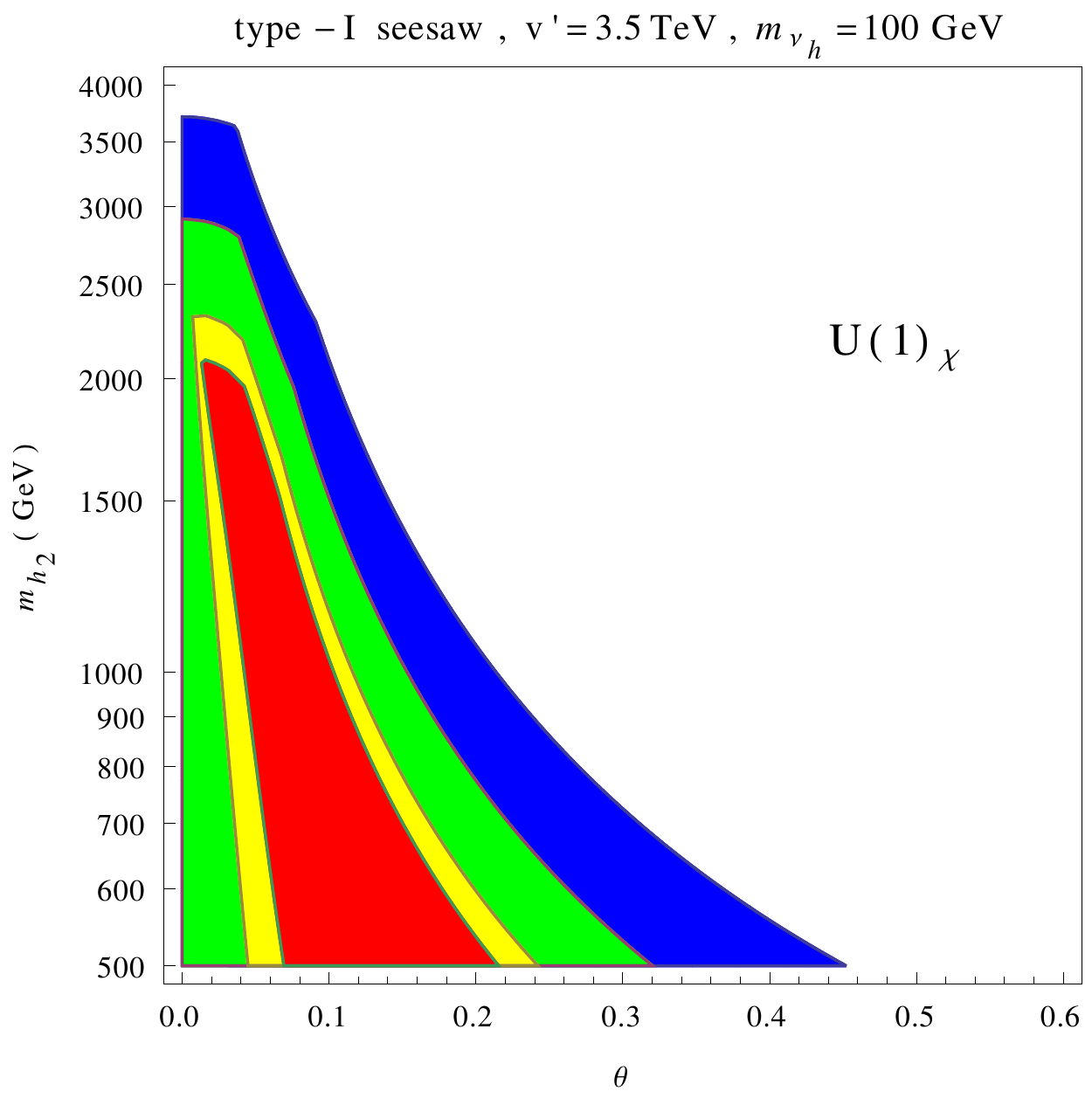}} 
\caption{Regions in the $(m_{h_2},\theta)$ parameter space in which the stability conditions are preserved up to $10^5\, \GeV$ (blue), $10^{9} \, \GeV$ (green), $10^{15} \, \GeV$ (yellow) and $10^{19} \, \GeV$ (red) in the $U(1)_{B-L}$ (a), $U(1)_R$ (b) and $U(1)_\chi$ (c) in the type-I seesaw scenario with $v'=3.5\,\textrm{TeV}$ and $m_{\nu_h} = 100\,\GeV$. \label{Fig.typeImh2th}}
\end{figure}
\begin{figure}[t]
\centering
\subfigure[]{\includegraphics[scale=0.45]{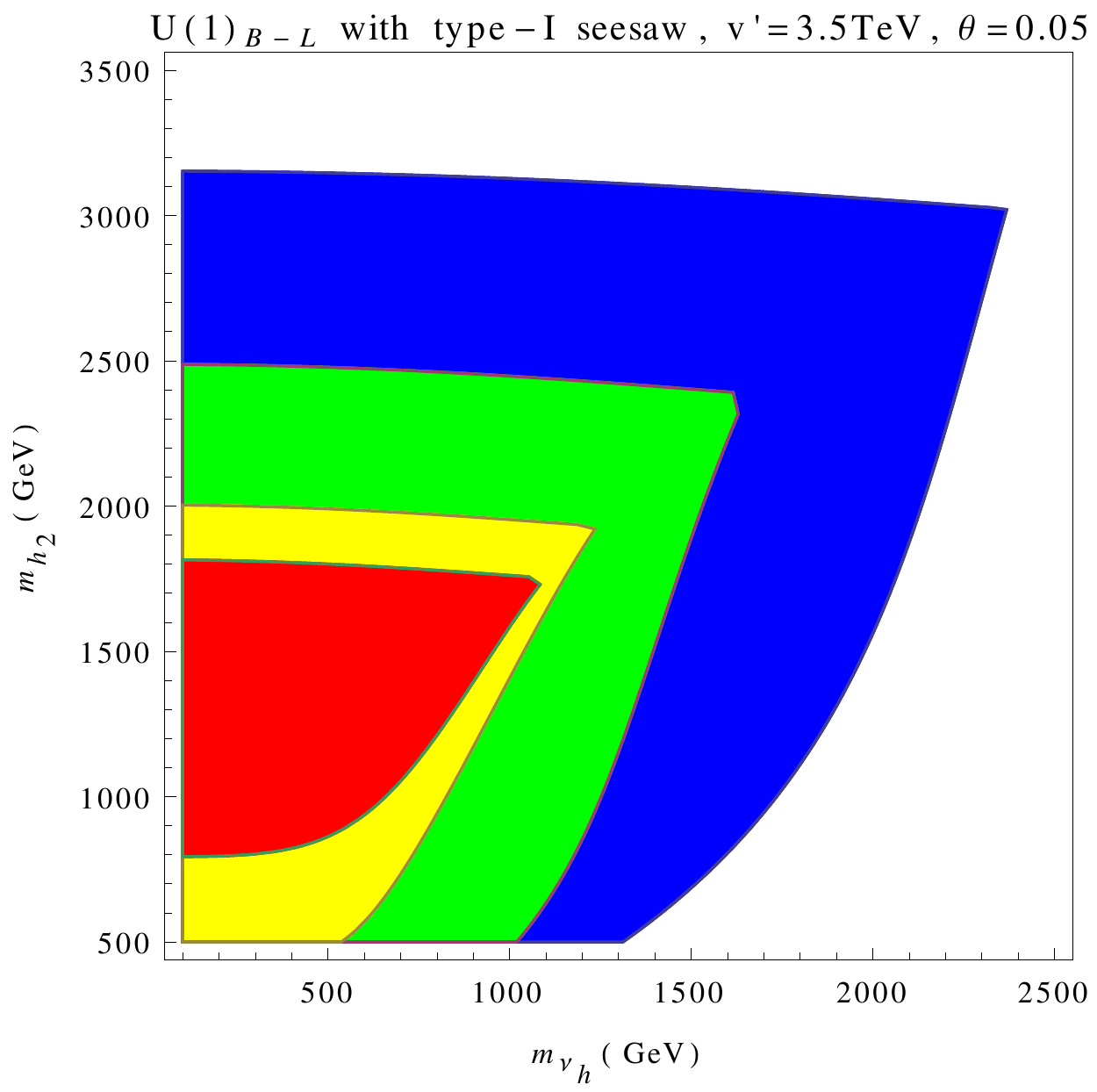}}
\subfigure[]{\includegraphics[scale=0.45]{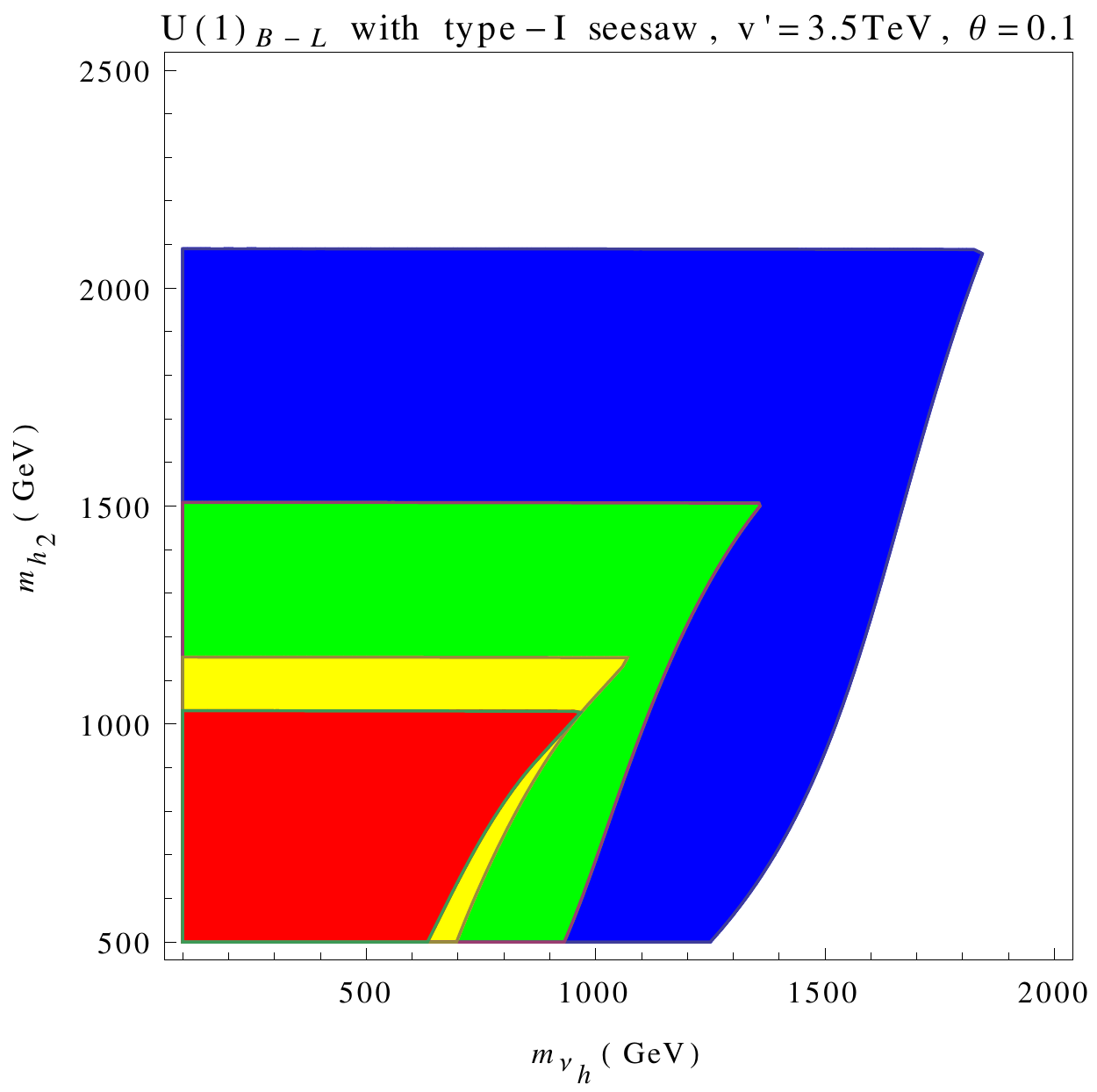}} 
\subfigure[]{\includegraphics[scale=0.45]{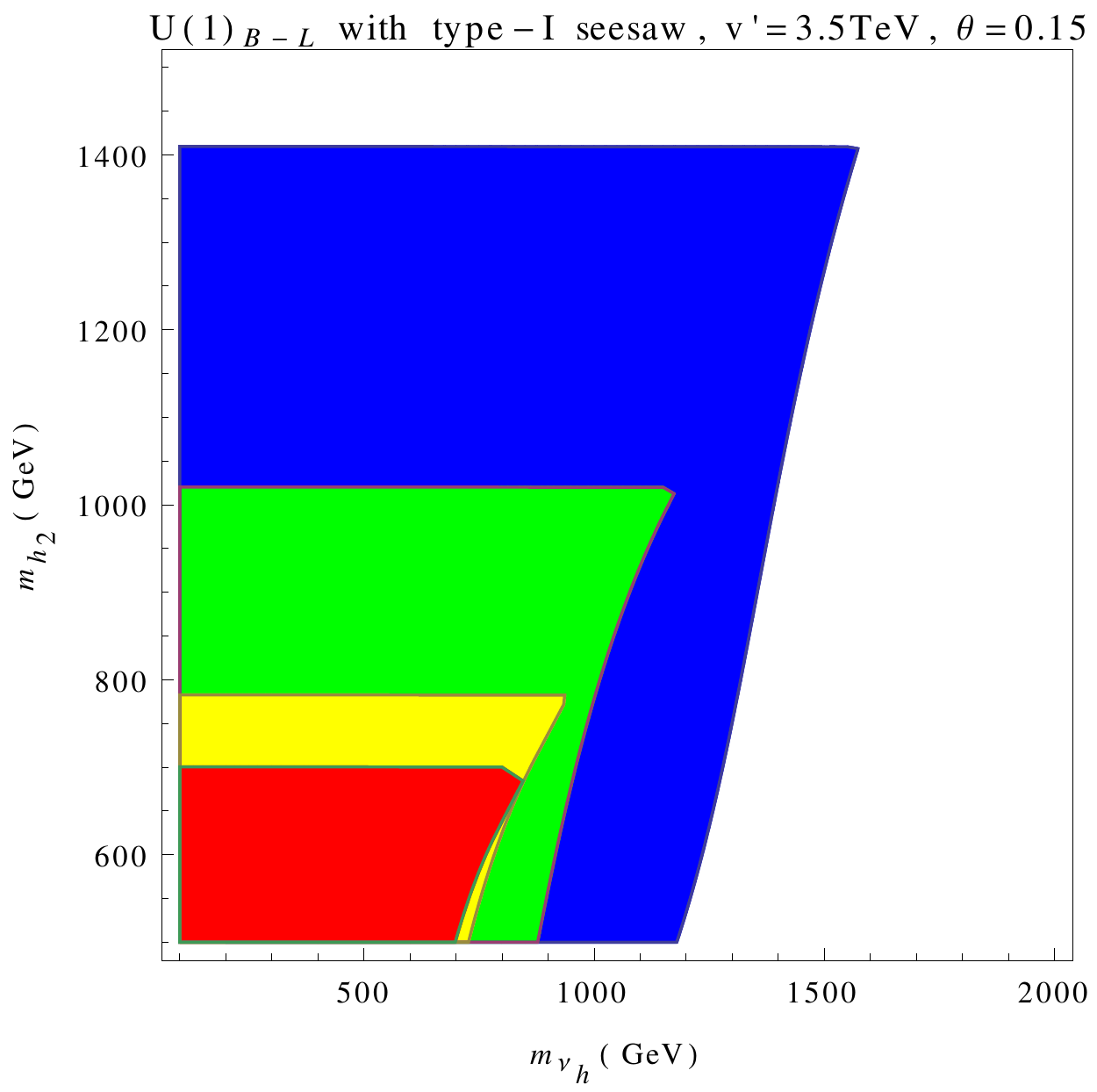}}  
\caption{Regions of the $(m_{h_2},m_{\nu_h})$ plane where the stability is preserved up to $10^5\, \GeV$ (blue), $10^{9} \, \GeV$ (green), $10^{15} \, \GeV$ (yellow) and $10^{19} \, \GeV$ (red) for a $U(1)_{B-L}$ extension. We have chosen $v'=3.5 \textrm{TeV}$ and $\theta = 0.05,0.1,0.15$. \label{Fig.typeImh2mnu}}.
\end{figure}
\begin{figure}[t]
\centering
\subfigure[]{\includegraphics[scale=0.6]{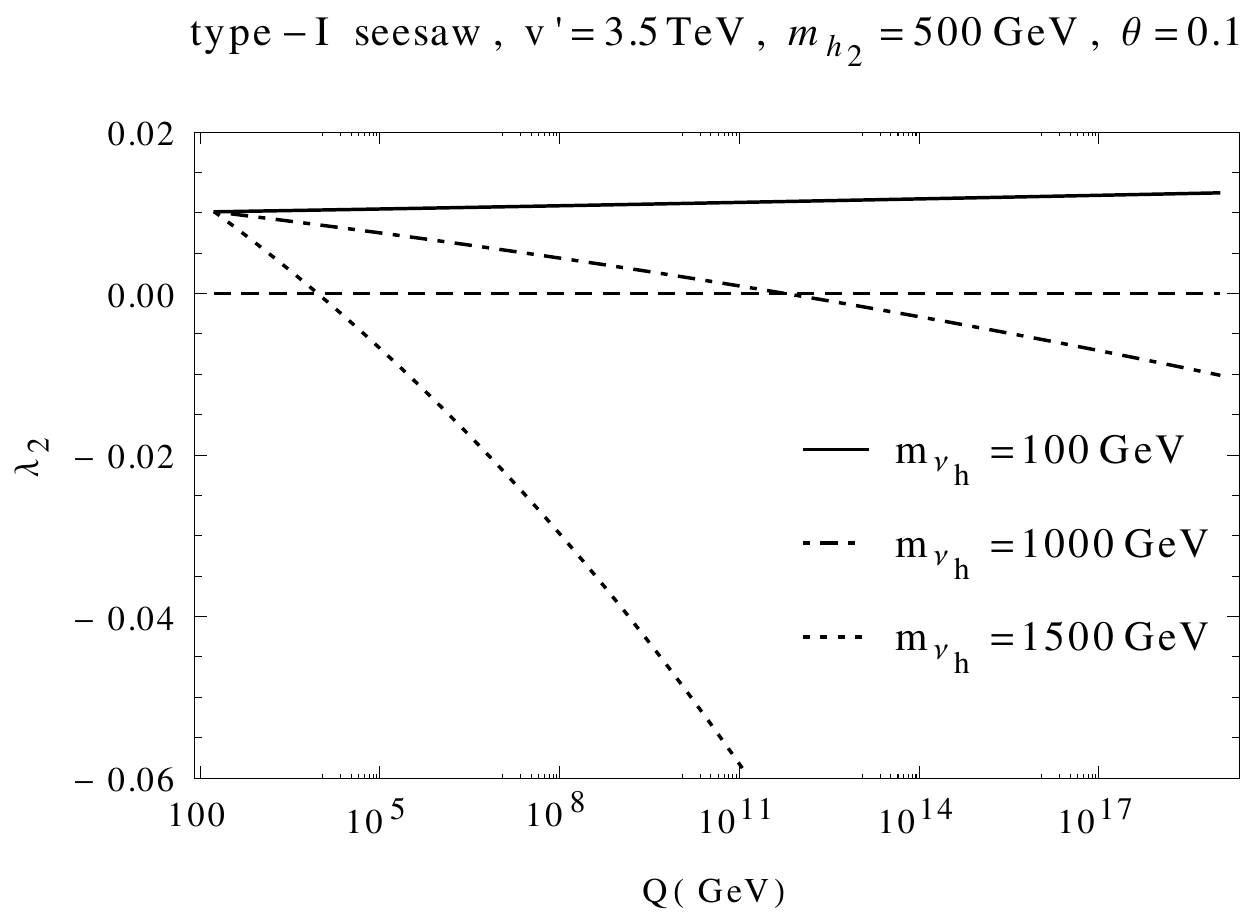}} \hspace{0.7cm}
\subfigure[]{\includegraphics[scale=0.6]{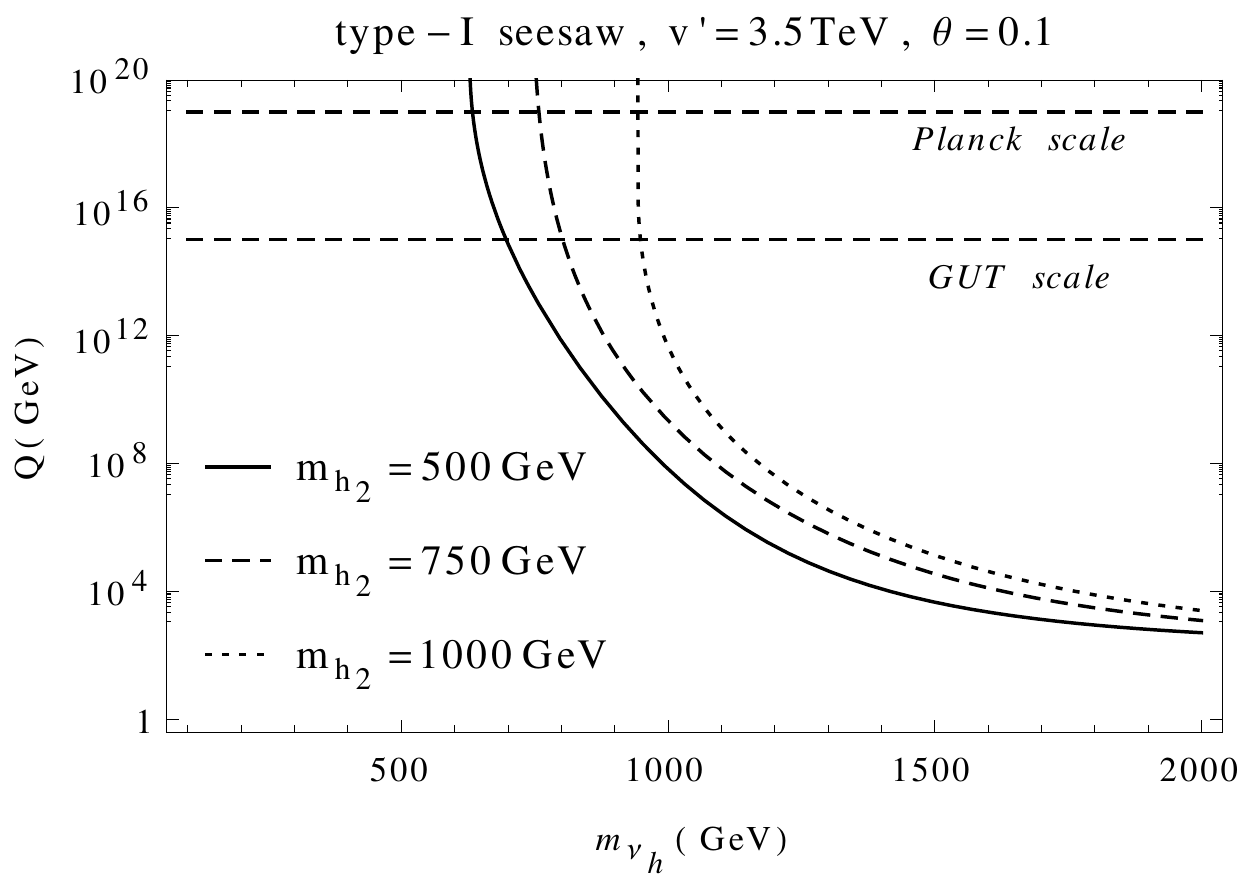}} 
\caption{(a) The evolution of $\lambda_2$ for different values of the heavy neutrino mass $m_{\nu_h}$ with $m_{h_2} = 500 \,\GeV$.
(b) The maximum scale up to which the stability conditions are fulfilled as a function of the heavy neutrino mass $m_{\nu_h}$.
The type-I seesaw in the $U(1)_{B-L}$ extension is considered, with $v'=3.5 \textrm{TeV}$ and $\theta = 0.1$. \label{Fig.typeIL2}}
\end{figure}
\begin{figure}[t]
\centering
\subfigure[]{\includegraphics[scale=0.45]{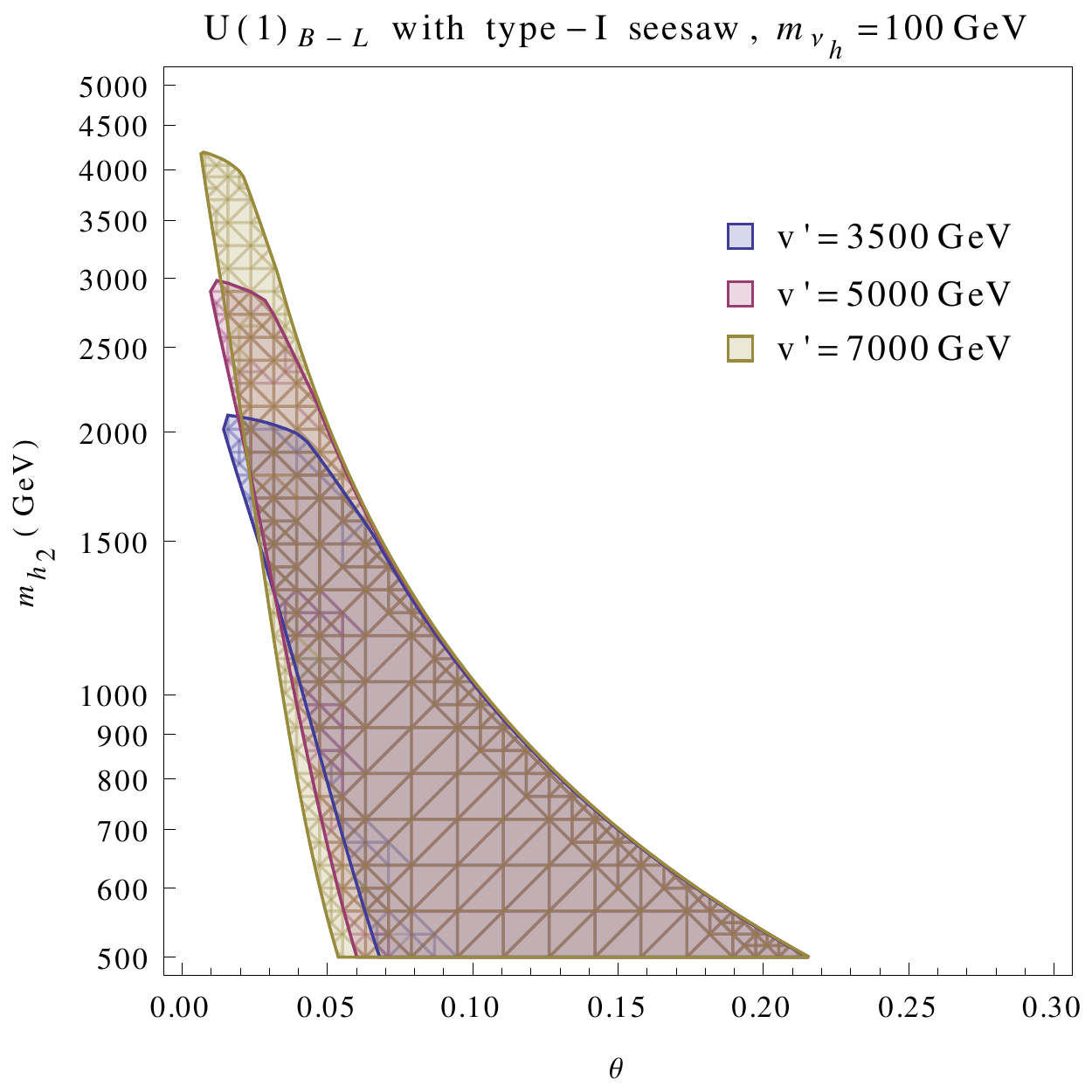}} \hspace{.7cm}
\subfigure[]{\includegraphics[scale=0.6]{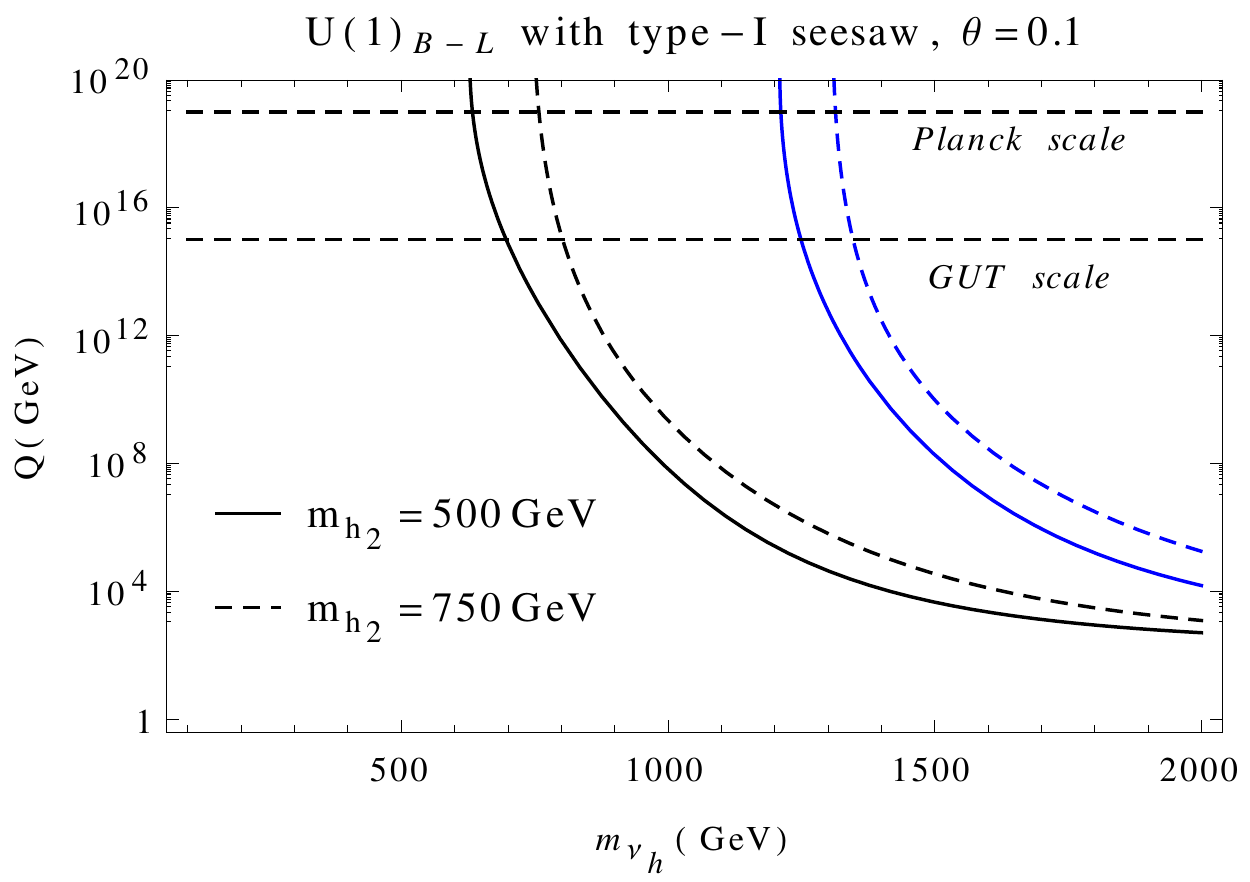}}
\caption{(a) Stability regions, up to the Planck scale, in a $U(1)_{B-L}$ extension with type-I seesaw, for three different vacuum expectation values of the heavy Higgs. (b) The maximum scale up to which the stability conditions are achieved as a function of $m_{\nu_h}$. The black (blue) lines correspond to $v'=3.5\,\textrm{TeV}$ ($v'=7 \, \textrm{TeV}$). \label{Fig.typeImh2vev}}
\end{figure}

\subsection{The stability bounds}
In the SM the analysis of the vacuum stability of the scalar potential is particularly simple and coincides with the requirement of the positivity of the only quartic coupling of the model. In a more complex model with two or more scalars, the full structure of the interaction potential must be taken into account, which generally involves nontrivial relations among the quartic coefficients. In our case, for instance, one has to study the constraints on the three couplings $\lambda_{1,2,3}$ given in Eq.~(\ref{stabilitycond}). Their values at the electroweak scale are deduced using Eq.~(\ref{lambdas}) in terms of the physical scalar masses, the two vacuum expectation values and the mixing parameter $\theta$. \\ 
In the following analyses we fix the mass of the light scalar $m_{h_1}$ at the SM Higgs mass of 125 GeV. In the SM Yukawa sector we retain only the top quark. Moreover, we use $g'=0.1$ and $\tilde{g} = 0$ at the electroweak scale. 
The free parameters of our models are, therefore, the mass of the heavy scalar $m_{h_2}$, the vacuum expectation value $v'$ of the SM singlet scalar, the scalar sector mixing angle $\theta$ and the Yukawa coupling $Y_N$. \\
A lower limit on the $v'$ vev can be deduced from the constraint
\bea
M_{Z'}/g' \ge 7 \, \textrm{TeV}
\eea
obtained by LEP-II at 99$\%$ C.L. \cite{Cacciapaglia:2006pk}. Indeed, assuming no-mixing in the neutral boson sector, $\tilde{g} = 0$, the $Z'$ mass is simply given by $M_{Z'} = |z_\phi| g' v'$. Therefore, in our case, using the fact that in all the three models that we investigate $|z_\phi|=2$, we adopt the lower bound $v'\ge 3.5\, \textrm{TeV}$. \\
The mixing in the scalar sector modifies the light scalar couplings to fermions and bosons with respect to the SM Higgs by a factor $\cos \theta$. Therefore the electroweak precision measurements can be used to constrain the scalar-mixing angle $\theta$ through the $S, T, U$ parameters \cite{Dawson:2009yx}, obtaining, for a $m_{h_1} = 125 \, \GeV$,
\bea
\label{STUbound}
\theta \lesssim 0.44 \qquad \mbox{with} \quad m_{h_2} \ge 500 \, \GeV \,.
\eea  
In the type-I seesaw scenario, the smallness of the light neutrino masses implies, as we have already seen, $Y_\nu \lesssim 10^{-6}$ which is too small to affect the RG equations of the scalar potential. This situation is typical of models with a type-I seesaw, and it is not shared by other cases, for instance by models with an inverse-seesaw mechanism where neither $Y_\nu$ nor $Y_N$ are constrained to small values. Regarding the heavy neutrinos, for the sake of simplicity, we assume their masses degenerate in flavour space, $m_{\nu_h} \equiv m_{\nu_h}^{1,2,3}$. This assumption simplifies the structure of the Yukawa coupling $Y_N$ which becomes proportional to the unit matrix. \\
We are now going to analyze the RG evolution of the scalar sector imposing the vacuum stability conditions defined in Eq.~(\ref{stabilitycond}) together with the triviality constraints on the quartic scalar couplings and on the abelian gauge couplings given in Eq.~(\ref{pertconds}). 
\subsubsection{Results }
In Fig.~\ref{Fig.typeImh2th} we show the stability regions in the $(m_{h_2},\theta)$ parameter space. In this case the heavy neutrino masses have been fixed to $m_{\nu_h} = 100 \, \GeV$. As usual the blue, green, yellow and red regions indicate the parameter space in which the stability condition is maintained up to $10^5\, \GeV$, $10^{9} \, \GeV$, $10^{15} \, \GeV$ and $10^{19} \, \GeV$ respectively. As we have already mentioned in the previous sections, the $Y_\nu$ Yukawa coupling is too small to affect the RG equations and can be neglected. Moreover, for a heavy neutrino mass $m_{\nu_h} \sim 100 \, \GeV$, $Y_N$ is also negligible. For a value of the mass of the heavy Higgs $m_{h_2} \lesssim 2 \,\textrm{TeV}$, the stability regions of the potential are found to cover an interval characterized by small values of the mixing angle $\theta$. Interestingly, the interval  overlaps with the one predicted by the bound given in Eq.~(\ref{STUbound}). This suggests that the smallness of the mixing between the two scalars can be inferred also from RG stability arguments, beside the indications coming from electroweak precision data. Such values of the mass of the heavy Higgs are necessary in order to achieve vacuum stability up to the GUT scale.  
From Fig. \ref{Fig.typeImh2th} it is also evident that the three charge assignments considered here do not provide very different results. Indeed, the shape of the allowed regions remain qualitatively the same. The only major difference between the three SM extensions analyzed here is in the $U(1)_R$ model, in which the stability regions up to $10^{15} \, \GeV$ and $10^{19} \, \GeV$ appear to be enlarged, with respect to the other two cases, and select smaller values of the mixing angle $\theta$. This behaviour shows that the effects of the charge assignments, which are in general quite small, can nevertheless influence the vacuum stability in specific regions of the parameter space. 

In Fig. \ref{Fig.typeImh2mnu} we investigate the stability regions in the $(m_{h_2}, m_{\nu_h})$ parameter space. Stability of the potential along the RG evolution reaches the Planck scale even for a heavy neutrino up to $m_{\nu_h} \sim 1\, \textrm{TeV}$, for $v'=3.5 \textrm{TeV}$. We observe that the heavy neutrino mass does not influence the upper bound on the heavy scalar mass, which essentially originates for the choice of $v'$ and $\theta$, as one can deduce from Fig. \ref{Fig.typeImh2th}. 

As we raise the masses of the heavy neutrinos, the effect of $Y_N$ quickly overcome all the other contributions in the running of $\lambda_2$, driving it towards negative values and, therefore, compromising the stability of the scalar sector. This behaviour is due to the large and negative $Y_N^4$ term in the $\beta$ function of $\lambda_2$ and can be exploited to extract upper bounds on the heavy neutrino masses. Fig. \ref{Fig.typeIL2}(a) illustrates the evolution of $\lambda_2$ in the $U(1)_{B-L}$ extension for different values of the heavy neutrino mass, namely $m_{\nu_h} = 100, 1000, 1500 \, \GeV$, which correspond, respectively, to $Y_N \simeq 0.02, 0.2, 0.3$. This clearly shows that the $\lambda_2$ quartic coupling quickly turns to negative values as the neutrino mass increases. 
Indeed, a $m_{\nu_h} \sim 2 \, \textrm{TeV}$, with $v' \sim 3.5 \, \textrm{TeV}$, easily violates the vacuum stability requirement below the \textrm{TeV} scale. \\
It is then interesting to study the behaviour of the maximum scale, up to which the stability is maintained, as a function of $Y_N$ or, equivalently, of $m_{\nu_h}$. This is shown in Fig. \ref{Fig.typeIL2}(b) for different values of the heavy Higgs mass. One can deduce that the heavy neutrino mass cannot be larger than $600-1000\,\GeV$ in order to achieve the stability up to the GUT or Planck scales. Notice that there are no significant changes on the limiting value of $m_{\nu_h}$ between the two scales, due to the presence of an asymptote in the three curves, which shows a boundless stability. One can easily observe that the allowed region for $m_{\nu_h}$ increases for a bigger $m_{h_2}$. Indeed, a heavier $h_2$ implies larger values for the quartic couplings at the electroweak scale, which compensate, at least in the range $600-1000\,\GeV$, the decreasing effect of a bigger $m_{\nu_h}$.

We conclude our analysis with some comments on the implications for the stability of the potential under a change in the vev of the extra SM singlet scalar $\phi$,  $v'$. In Fig. \ref{Fig.typeImh2vev} (a) we have depicted the regions in the $(m_{h_2},\theta)$ space in which stability is achieved up to the Planck scale for three different values of the $v'$ vacuum expectation value. It is interesting to observed that a bigger $v'$ only affects the small $\theta$ region, extending the maximum allowed values of the heavy scalar mass. On the other hand, as one can see from Fig. \ref{Fig.typeImh2vev}(b), the stability bounds on the heavy neutrino masses trivially scale with $v'$. Indeed black curves correspond to $v'=3.5\,\textrm{TeV}$ while the blue ones are obtained for $v'=7\,\textrm{TeV}$. This is due to the fact that the RG equations depend explicitly on the dimensionless $Y_N$ and $Y_N = m_{\nu_h}/(\sqrt{2} v')$. \\
If $v'$ is pushed well above the TeV scale, the SM particles would decouple from the new degrees of freedom introduced by the addition of the extra abelian gauge group and of the heavy scalar field. In such a case the evolution of the parameters of the scalar sector are controlled by the SM particle content alone only up to the $U(1)_{B-L}$ spontaneous symmetry breaking scale. Only at this scale the complete model should be taken into account, and the evolution would be driven by the $\beta$ functions presented in Eq.(\ref{betafuncs}). In this case, as pointed out in \cite{EliasMiro:2012ay}, the tree-level threshold corrections in the scalar sector have to be included and this would help in stabilizing the scalar potential. On the other hand, as $v'$ reaches higher scales, it would be possible to reproduce the small neutrino masses even with $Y_\nu \sim O(1)$. In this situation, the larger values of the Yukawa's of the light neutrinos would effect the RG evolution and should be taken into account. The analysis of this scenario is beyond the scope of the present paper and it will be discussed in a separate work.

\section{Conclusions} 
We have presented results of a numerical study of the RG equations for a generic $U(1)'$ extension of the SM with the inclusion of one extra complex scalar and three heavy right handed neutrinos. We have conisdered a type-I seesaw mechanism for the generation of the small masses of the three light SM neutrinos.
The $U(1)'$ charges have been determined by requiring the cancellation of the gauge and gravitational anomalies. Our work has been based on a re-analysis of the evolution, which is drastically affected by the coefficient of the Yukawa's of the right handed neutrinos. As we have pointed out in the introduction, a crucial change in the RG equation for the coefficient of the quartic coupling $\lambda_2$ makes such a coupling negative even at the TeV scale, destabilizing the potential.

We have discussed, in the 2-parameter class of solutions that we have investigated, three specific charge assignments, corresponding to $U(1)_{B-L}, U(1)_R$ and $U(1)_\chi$, showing that their RG evolutions 
share similar behaviour.  We have also focused our interest on a specific scenario in which the vev of the extra Higgs is in the TeV scale, which could be studied at the LHC. We have shown that within this scenario, for a heavy Higgs in the TeV range and a right handed neutrino of mass between 100 and 1000 GeV, all the constraints coming from the vacuum stability of the scalar potential are satisfied. For this reason, larger mass values of the right handed neutrino do not allow to extend the validity of these models up to the Planck scale. More details of this analysis will be presented in a forthcoming work.

\centerline{\bf Acknowledgments} 
We thank Lorenzo Basso for correspondence.

\end{document}